\documentclass[a4paper,11pt]{article}
\pdfoutput=1 

\usepackage{jheppub} 

\usepackage[T1]{fontenc} 

\title{\boldmath Frenkel electron and a spinning body in a curved background}

\author[a,1]{Walberto Guzm\'an Ram\'irez, \note{Corresponding author.}}
\author[a]{Alexei A. Deriglazov,}
\author[a]{Andrey M. Pupasov-Maksimov}

\affiliation[a]{Depto. de Matem\'atica, ICE, Universidade Federal de Juiz de Fora, MG, Brasil}

\emailAdd{wguzman@cbpf.br}
\emailAdd{alexei.deriglazov@ufjf.edu.br}
\emailAdd{pupasov@phys.tsu.ru}

\abstract{We develop a variational formulation of a particle with spin in a curved space-time background. The model is based on a
singular Lagrangian which provides equations of motion,  a fixed value of spin and Frenkel condition on spin-tensor.
Comparing our equations with those of Papapetrou we conclude that the Frenkel electron in a gravitational field has the
same behavior as a rotating body in the pole-dipole and leading-spin approximation. Due to constraints presented in the
formulation, position space is endowed with a noncommutative structure induced by the spin of the particle. Therefore, the
model provides a physically interesting example of a noncommutative particle in a curved background.
}

\begin{document}
\maketitle
\flushbottom

\section{Introduction \label{intro}}
The behavior of a spin one-half elementary particle and a rotating body (spinning test particle) in an external
electromagnetic or gravitational background has long been a subject of intensive study in physics. In both cases,
non-relativistic theory can be described in terms of the center-of-mass and three-vector of spin ${\bf S}$ (for the case of
a body this happens in the pole-dipole approximation \cite{fock1939mouvement, Papapetrou:1951pa}). Spin represents the
internal angular momentum of a particle, its torque leads to precession. The first attempt to describe a spin one-half
particle in relativistic theory was due to Thomas \cite{Thomas}; he pointed out a relativistic correction to the spin
precession. Frenkel observed \cite{Frenkel} that the natural description of spin in special relativity can be achieved by 
including ${\bf S}$ in the antisymmetric spin-tensor $J^{\mu\nu}$, $J^{ij}=2\epsilon^{ijk}S^k$. In this case, the right
balance between non-relativistic and relativistic degrees of freedom is achieved by imposing the condition $J^{\mu\nu}\dot
x_\nu=0$. He also noticed that construction of a consistent system of equations (as well as of a variational problem)
for position and spin-tensor, subjected to this condition, represents a rather nontrivial problem.

To describe polarization effects of an electron in uniform electromagnetic fields, Barg\-mann,  Michael and Telegdi
(BMT) \cite{BMT} suggested equations of motion in terms of a four-vector of spin instead of a spin-tensor.
In the linear approximation on electromagnetic field strength, the BMT equations agree with those of Frenkel.

Various variational formulations for a spin one-half particle have been proposed\footnote{Note that one needs Hamiltonian
to apply Frenkel or BMT-theory for description of Zeeman effect.}, see \cite{corben:1968, hanson1974relativistic,
grassberger1978, cognola1981lagrangian, berezin:1977, Rietdijk, Ravndal, AAD5, Alexei,DPM3, AAD12} and references therein.
In the present work, we use the unified formulation for the Frenkel and BMT equations developed in \cite{Alexei}. The
model is based on a singular Lagrangian which implies all necessary conditions on spin in a course of the Dirac-Bergmann
algorithm. Besides the position coordinate $x^\mu$ and its momentum $p^\mu$, our formulation involves an extra vector-like
variable $\omega^\mu$ with conjugated momentum $\pi^\mu$. These can be used to construct both the Frenkel-type spin-tensor
and BMT-vector. The Frenkel condition is guaranteed by two constraints\footnote{$p^\mu\sim\dot x^\mu$ in free theory.
In interacting theory the condition $S^\mu p_\mu=0$ on BMT spin-vector implies $S^\mu\dot x_\mu=0$, see Eq.
(\ref{uf4.121}).} appeared for basic variables, $p\omega=p\pi=0$. The  essential point is that the second-class
constraints must be taken into account by transition from Poisson to Dirac brackets. As the constraints involve
conjugate momenta $p^\mu$ for $x^\mu$, this leads to nonvanishing brackets for the position variables. As a result,
the position space is endowed, in a natural way, with  a noncommutative structure which originates from taking into account the spin
degrees of freedom. Hence our spinning particle provides an example of noncommutative relativistic system, with
noncommutative geometry of position space induced by the spin of the particle. Canonical quantization of the particle gives
\cite{DPM1} the one-particle sector of the Dirac equation, so we refer to the particle as Frenkel electron.

The conventional equations of a rotating body  in gravitational field (in pole-dipole approximation) were formulated
from the analysis of compatibility condition of the energy-momentum tensor \cite{fock1939mouvement, Papapetrou:1951pa,
Mathisson:1937zz}. These equations, referred to as the Mathisson-Papapetrou (MP) equations, do not form a closed system
\cite{Papapetrou:1951pa}. Their closure requires an auxiliary Frenkel-type condition on the spin-tensor. Another drawback
is that the position coordinate obeys a  third-order differential equation \cite{Papapetrou:1951pa, Pomeranskii}.
Nevertheless, since the derivation of the MP equations, they have been applied in different scenarios of general
relativity including Schwarzschild and Kerr backgrounds, gravitational waves \cite{Corinaldesi:1951pb} etc. Influence
of a curved background on the motion of a gyroscope allows one to test general relativity in a laboratory (nuclear spin
gyroscope measurements of the Earth rotation \cite{kornack2005nuclear-gyroscope}) or orbital experiments (Probe B
experiment). Such an influence can be more important for the relativistic regimes \cite{chicone2005relativistic}.

One of the advantages of a variational formulation is in consistent theoretical definition of all variables of a model.
Different variational formulations for spinning test particles in gravitational fields exist \cite{Hojman, Porto, Karl,
Burdet}. In \cite{Hojman, Porto} the spin-tensor is defined as the conjugated momentum to the angular velocity tensor
which is a combination of the tetrad fields and their derivatives. The action is constructed from all invariant
quantities that can be formed using the velocity vector and the angular-velocity tensor. In the cited works,
supplementary conditions on spin should be added by hands to ensure the correct number of degrees of freedom. In
\cite{Karl},  the authors used the formalism of Routhian to reach the MP equations.


The aim of the present work is a two-fold. First, we  formulate the variational problem for the Frenkel electron in an
arbitrary curved space-time background and study the corresponding equations of motion. Second, we show that in an
arbitrary background our equations are compatible with the  MP equations, and coincide with them in the weak-field
approximation. In other words, the Frenkel electron in a gravitational field has the same behavior as a rotating body in the 
pole-dipole approximation.

The work is organized as follows. To make it self-contained, in section 2 we give  a short review of the Frenkel and BMT
approaches and describe the variational formulation of the Frenkel electron in the presence of an external electromagnetic
field. In section 3 we develop a variational formulation of the Frenkel electron in a curved background and study both
the Hamiltonian and Lagrangian equations of motion. In section 4 we consider the weak-field limit and compare this with
other approaches.


\section{Variational problem for Frenkel and Bargmann-Michel-Telegdi equations \label{subsec:review-BMT}}


Frenkel \cite{Frenkel} has included the three-dimensional spin-vector $S_i$ into the antisymmetric tensor
$J^{\mu\nu}=-J^{\nu\mu}$ obeying the constraint
\begin{eqnarray}\label{intr.13}
J^{\mu\nu}u_\nu=0,
\end{eqnarray}
where $u_\nu=\dot x_\nu(s)$ represents four-velocity of particle in the proper-time parametrization $s$, $\dot
x^2(s)=-c^2$. In the rest-frame, $u_\nu=(u_0, 0, 0, 0)$, this implies $J^{0i}=0$. So only three components
$J^{ij}=2\epsilon^{ijk}S^k$ of the Frenkel tensor survive. Besides, we can impose the scalar constraint
\begin{eqnarray}\label{intr.14}
J^{\mu\nu}J_{\mu\nu}=6\hbar^2.
\end{eqnarray}
As in the rest frame we have $J^{\mu\nu}J_{\mu\nu}=8(S^i)^2$, this implies that we work with spin-$1/2$ particle.

Frenkel tensor is equivalent to the four-vector
\begin{equation}
s^\mu\equiv\frac{1}{4\sqrt{-u^2}}\epsilon^{\mu\nu\alpha\beta}u_\nu J_{\alpha\beta},
\end{equation}
%
%
%
which has been taken by Bargmann, Michel and Telegdi as the basic quantity in their description of spin.  The
BMT-vector of spin obeys the following  constraints
\begin{eqnarray}\label{intr.16}
s^\mu u_\mu=0, \quad (s^\mu)^2=\frac{3\hbar^2}{4}.
\end{eqnarray}
Starting from the non-relativistic precession equation in the rest frame and imposing compatibility with above
mentioned constraints we obtain BMT-equations \cite{BMT}
\begin{eqnarray}
\ddot x^\mu&=&\frac{e}{mc}(F\dot x)^\mu, \label{intr.17} \\
\dot s^\mu&=&\frac{e\mu}{mc}(Fs)^\mu-\frac{e}{mc^3}(\mu-1)(sF\dot x)\dot x^\mu\,, \label{intr.18}
\end{eqnarray}
for a particle with mass $m$, charge $e$ and magnetic moment $\mu$. In terms of $J_{\mu\nu}$ equation (\ref{intr.18})
reads
\begin{equation}\label{pp22-J}
\dot J^{\mu\nu}=\frac{e}{mc} \left[\mu F^{[\mu}{}_\alpha J^{\alpha\nu]}- \frac{(\mu-1)}{c^2}(JF\dot x)^{[\mu}\dot
x^{\nu]}\right].
\end{equation}

Let us shortly describe unified variational formulation for both Frenkel and BMT equations \cite{Alexei, DPM3}. The
basic variables of the model are position $x^\mu(\tau)$ and a vector $\omega^\mu(\tau)$ taken in arbitrary
parametrization $\tau$. Besides, Lagrangian depends on auxiliary variables $e_i(\tau)$, $i=1, 3, 4, 7$. At the end,
they provide all the necessary constraints. Conjugate momenta are denoted as $p^\mu$, $\pi^\mu$ and $\pi_{ei}$.
Consider the following action
\begin{eqnarray}\label{lagrangian-bmt-em-4aux-vars}
S=  \int d\tau\quad \frac{1}{2\Delta}\left[ e_3\left(\dot xN\dot x\right)-2e_7\left(\dot xND\omega\right)
+e_1\left(D\omega ND\omega\right) \right] +\frac{e}{c}A_\mu\dot x^\mu- \cr
\frac{e_4}{2}(\omega^2-a_4)-\frac{e_1}{2}m^2c^2+\frac{e_3}{2}a_3 \, ,
\end{eqnarray}
which depends on two numeric parameters $a_3$ and $a_4$. It has been denoted
%
$\Delta = e_1e_3 -e_7^2$, $N^{\mu\nu}\equiv \eta^{\mu\nu} - \frac{\omega^\mu\omega^\nu}{\omega^2}$.
%
$F_{\mu\nu}$ is the electromagnetic tensor and $A_\mu$ its vector potential. The Lagrangian contains
a non-minimal interaction
\begin{eqnarray}\label{il1}
D\omega^\mu=\dot\omega^\mu-e_1\frac{e\mu}{c}(F \omega)^\mu,
\end{eqnarray}
of spin variable $\omega^\mu$  with the electromagnetic field. There is no translation symmetry in the space
$\omega^\mu$, so $\omega^\mu$ represents a vector-like variable. According to normalization constraint $\omega^2=a_4$,
the ends of vectors lie on a hyperboloid. We could resolve the constraint and then to omit the term
$\frac{e_4}{2}(\omega^2-a_4)$ from the action, but this lead to lose of manifest relativistic covariance of the
formalism. Besides, we point out that the vectors $\omega^\mu$ change under spin-plane local symmetry. For detailed
analysis of resulting phase-space geometry see \cite{DPM2}. In short, in non-relativistic models of spin, inner
manifold which describes the spin degrees of freedom is two sphere. The constraint $\omega^2=a_4$ represents a
covariant generalization of this construction. It can be seen by transition to the rest frame for the free particle. In
this frame $\omega^0=0$, and spacial components $\vec{\omega}$ belong to the sphere $\vec{\omega}^2=a_4$.

The action is invariant under two-parametric group of local symmetries which consist of reparametrizations and
spin-plane transformations, see \cite{deriglazov2012variational} for details. As we deal with a singular Lagrangian,
the Dirac-Bergmann algorithm is necessary to study the model in Hamiltonian formulation. Performing transition to the
Hamiltonian formulation, we find the following constraints
\begin{equation}\label{intr.22}
{\cal{P}}^2-\frac{\mu e}{2c}F^{\mu\nu}J_{\mu\nu}+m^2c^2=0,
\end{equation}
\begin{eqnarray}
\pi^2-a_3=0, \qquad \omega^2-a_4=0, \qquad  \omega\pi=0, \label{intr.20} \\
{\cal P}\omega=0, \qquad {\cal P}\pi=0. \label{intr.21}
\end{eqnarray}
The equation (\ref{intr.22}) is the standard mass-shell condition \cite{Holten, Nash}. This turns out to be the
first-class constraint. ${\cal P_\mu}$ is defined in terms of conjugate momenta $p_\mu$ and the electromagnetic vector
potential as follows
\begin{equation}\label{il3}
{\cal P}^\mu=p^\mu-\frac{e}{c}A^\mu\,.
\end{equation}
The relation with Frenkel theory is reached by defining the Frenkel-type spin-tensor
\begin{eqnarray}\label{intr.19}
J^{\mu\nu}=2(\omega^\mu\pi^\nu-\omega^\nu\pi^\mu).
\end{eqnarray}
Then Eqs. (\ref{intr.21}) guarantee its
transversality
\begin{equation}
J^{\mu\nu}{\cal P}_\nu=0,
\end{equation}
while Eqs. (\ref{intr.20}) tell us that
\begin{equation}
J^{\mu\nu}J_{\mu\nu}=8(\omega^2\pi^2-(\omega\pi)^2)=8a_3a_4.
\end{equation}
This fixes the value of spin by an appropriate choice of the numbers $a_3$ and $a_4$ .

The set (\ref{intr.20}), (\ref{intr.21}) contains one first-class constraint.  Taking into account that each
second-class constraint rules out one phase-space variable, whereas each first-class constraint rules out two
variables, we have the right number of spin degrees of freedom, $8-(4+2)=2$.

For the case of an arbitrary background, the action leads to rather complicated equations of motion, see \cite{DPM3}.
So we present here their form in proper-time parametrization, for uniform electromagnetic field
\begin{eqnarray}\label{pp20}
\frac{d(\tilde T \dot x)^\mu}{d\tau}=\frac{e}{m_rc}(F\dot x)^\mu\,,
\end{eqnarray}
\begin{equation}
\label{pp19-J-F0} \dot J^{\mu\nu} =\frac{e}{m_rc} \left[\mu F^{[\mu}{}_\alpha J^{\alpha\nu]}-
\frac{m_r^2(\mu-1)}{c^2M^2}(JF\tilde T\dot x)^{[\mu}(\tilde T\dot x)^{\nu]}\right]\,,
\end{equation}
\begin{eqnarray}\label{pp21}
\dot s^\mu=\frac{e\mu}{m_rc}(Fs)^\mu-\frac{e}{m_rc^3}\left[(\mu-1)(sF\dot x)+\mu b(sFJF\dot x)\right](\tilde T\dot
x)^\mu\,.
\end{eqnarray}
The following notation were used
\begin{equation}
\tilde T^{\mu\nu}\equiv \eta^{\mu\nu}-\frac{2e(\mu-1)}{2m^2c^3(3\Delta^2-1)}(JF)^{\mu\nu}, \quad \Delta^2\equiv
1-\frac{\mu e}{2m^2c^3}(JF),
\end{equation}
\begin{eqnarray}\label{il0}
m_r^2=m^2-\frac{\mu e}{2c^3}(JF)\,, \qquad M^2=m^2-\frac{e(2\mu+1)}{4c^3}(JF).
\end{eqnarray}
These equations are written in an arbitrary parametrization of the world-line.  In (\ref{pp21}) the spin vector if defined by
\begin{equation}\label{def:4-vector-of-spin}
s^\mu\equiv\frac{1}{4\sqrt{-{\cal P}^2}}\epsilon^{\mu\nu\alpha\beta}{\cal P}_\nu J_{\alpha\beta}.
\end{equation}
By construction, $s^\mu$ obeys
\begin{eqnarray}\label{uf4.121}
s^\mu{\cal P}_\mu=s^\mu\dot x_\mu=0\,, \qquad s_\mu s^\mu=a_3a_4\,.
\end{eqnarray}
The BMT equations (\ref{intr.17}) and (\ref{pp22-J}) appeared neglecting higher-order terms of $\hbar$ in equations
(\ref{pp20}) and (\ref{pp21}). In the case of a free particle, $F^{\mu\nu}=0$, both spin-tensor and spin-vector are
conserved quantities, $\dot J^{\mu\nu}=0$, ~ $\dot s^\mu=0$.


\section{Spinning particle in a curved space-time \label{sec:curved-ST}}
We obtain variational formulation of Frenkel particle in a curved background by writing the action
(\ref{lagrangian-bmt-em-4aux-vars}) with $A^\mu=0$ in a generally covariant form. It is sufficient to replace the
Minkowski metric $\eta_{\mu\nu}$ by an arbitrary metric $g_{\mu\nu}(x^\rho)$, and usual derivative of contravariant
vector $\omega^\mu$ by the covariant derivative
\begin{equation}\label{s-g}
\dot\omega^\mu ~ \rightarrow ~ D\omega^\mu=\dot\omega^\mu + \Gamma^\mu_{\alpha\beta}\dot x^\alpha\omega^\beta.
\end{equation}
%
Then the action reads
\begin{eqnarray}\label{L-curved}
S_g= \int d\tau \quad \frac{1}{2\Delta}\left[ e_3\left(\dot xN\dot x\right)-2e_7\left(\dot xND\omega\right)
+e_1\left(D\omega ND\omega\right) \right]- \frac{e_4}{2}(\omega^2-a_4)- \cr \frac{e_1}{2}m^2c^2 +\frac{e_3}{2}a_3 \, .
\end{eqnarray}
Velocities $\dot x^\mu$, $D\omega^\mu$ and projector $N^{\mu\nu}=g^{\mu\nu}-\frac{\omega^\mu\omega^\nu}{\omega^2}$
transform as contravariant tensors, therefore the
Lagrangian is manifestly invariant under general coordinate transformations. In the flat limit,
$g_{\mu\nu}=\eta_{\mu\nu}$, the action (\ref{L-curved})  reduces to  (\ref
{lagrangian-bmt-em-4aux-vars}) with $A^\mu=0$.


\subsection{Hamiltonian constraints}
Let us construct Hamiltonian formulation of the theory (\ref{L-curved}).
The conjugate momenta of $\omega^\mu$ and $x^\mu$ are
%
\begin{eqnarray}
\pi_\mu&=&-\frac{e_7}{\Delta}N_{\mu\nu} \dot x^\nu + \frac{e_1}{\Delta_e}N_{\mu\nu}D\omega^\nu\,, \label{pi-omega}\\
p_\mu&=&\frac{e_3}{\Delta}N_{\mu\nu}\dot x^\nu -\frac{e_7}{\Delta_g}N_{\mu\nu}D\omega^\nu +
\Gamma^\alpha_{\mu\nu}\omega^\nu\pi_\alpha\,. \label{p-x}
\end{eqnarray}
According to this expression, $\pi_\mu$ is a covariant vector.
%
%
Concerning the second line, due to the presence of Christoffel symbols, $p_\mu$ does not transform as a vector. Using
transformation law of $\Gamma^\alpha_{ \mu\nu}$ and the fact that other quantities transform as vectors, we get
transformation law of $p^\mu$
\begin{equation}\label{trans-p}
p'_\mu=\frac{\partial x^\alpha}{\partial x'^\mu} p_\alpha +\frac{\partial x'^\nu}{\partial x^\beta} \frac{\partial^2 x^\sigma}{\partial x'^\mu \partial
x'^\nu}\omega^\beta\pi_\sigma.
\end{equation}
Eq. (\ref{p-x}) prompts to introduce the covariant momentum $P_\mu$ which transforms as a vector
\begin{equation}\label{momenta-P}
P_\mu = p_\mu - \Gamma^\alpha_{\mu\beta} \omega^\beta\pi_\alpha.
\end{equation}
The projector $N^{\mu\nu}$ has  null vector $\omega_\mu$, therefore equations (\ref{pi-omega}) and (\ref{p-x}) imply
the primary constraints
\begin{eqnarray}\label{T_5andT_6}
T_5&\equiv &\omega^\mu \pi_\mu =0, \nonumber\\
T_6&\equiv& \omega^\mu P_\mu =0.
\end{eqnarray}
Conjugate momenta of $e_i$ vanish,  $\pi_{ei}=0$ , therefore we get four more primary constraints.
Applying the Legendre transformation to (\ref{L-curved}) and using (\ref{pi-omega}), (\ref{p-x}), we obtain the following Hamiltonian
\begin{eqnarray}\label{Hamiltonian-curved}
H=&&\frac{1}{2} e_1\left(g^{\mu\nu}P_\mu P_\nu + m^2c^2 \right) +\frac{1}{2}e_3\left(g^{\mu\nu}\pi_\mu\pi_\nu-a_3\right) +\\ \nonumber
&&+ \frac{1}{2}e_4\left( g_{\mu\nu}\omega^\mu \omega^\nu-a_4\right)+\lambda_5\pi_\mu \omega^\mu + e_7
g^{\mu\nu}P_\mu\pi_\nu + \lambda_6P_\mu\omega^\mu+\lambda_{ei}\pi_{ei},
\end{eqnarray}
where $\lambda_5$ and $\lambda_6$ appear as lagrangian multipliers for primary constraints $T_5$ and $T_6$, respectively,
and $\lambda_{ei}$ stand for Lagrangian multipliers associated with the primary constraints $\pi_{ei}$.
The Hamiltonian (\ref{Hamiltonian-curved}) consist of covariant quantities only, hence it is manifestly invariant under
general-coordinate transformations. Following the Dirac procedure for singular systems we conclude that
\begin{eqnarray}\label{constraints}
T_1 = g^{\mu \nu}P_\mu P_\nu + m^2c^2 =0,&&\quad T_3=g^{\mu\nu}\pi_\mu \pi_\nu - a_3 = 0, \nonumber \\
T_4=g_{\mu\nu}\omega^\mu \omega^\nu - a_4 =0,  &&\quad T_7=g^{\mu\nu}P_\mu \pi_\nu =0,
\end{eqnarray}
appear as secondary constraints when we impose the compatibility conditions $\dot\pi_{ei}=0$. The second, third and
fourth stages of the Dirac-Bergmann algorithm can be resumed as follows
\begin{eqnarray}\label{third-stage-L4toHam}
T_1=0 ~ \quad &\Rightarrow& \quad  \lambda_6R_{(\omega)}+e_7R_{(\pi)}=0\,,\label{il2.1} \\
T_3=0 ~ \quad &\Rightarrow& \quad ~  \lambda_5=0\,,\label{il2.2} \\
T_4=0 ~ \quad &\Rightarrow&\quad ~  \lambda_5=0\,,\label{il2.3} \\
T_5=0 ~ \quad &\Rightarrow&\quad ~ a_3e_3-a_4e_4=0\,, ~ \qquad \quad \Rightarrow\quad ~
\lambda_{e_4}=\frac{a_3}{a_4}\lambda_{e3}\,,\label{il2.4} \\
T_6=0 ~ \quad &\Rightarrow&\quad ~ e_1R_{(\omega)}+2e_7 M^2=0\,, \quad\Rightarrow\quad ~ \lambda_{e7}=f(\lambda_{e1})\,,\label{il2.5} \\
T_7=0 ~ \quad &\Rightarrow&\quad ~ e_1R_{(\pi)}-2\lambda_6 M^2=0\,, ~ \label{il2.6}
\end{eqnarray}
where we defined the scalar quantities
\begin{eqnarray}
R_{(\pi)}&\equiv& 2R_{\alpha\beta\mu\nu} \pi^\alpha \omega^\beta \pi^\mu P^\nu , \qquad R_{(\omega)}\equiv 2R_{\alpha
\beta\mu\nu} \pi^\alpha \omega^\beta \omega^\mu P^\nu  , \label{Rpi} \\
M^2&\equiv& m^2c^2 - R_{\alpha \mu\beta\nu}\omega^\alpha\pi^\mu \omega^\beta\pi^\nu. \label{byM.10}
\end{eqnarray}
Here $R_{\mu\nu\alpha\beta}$ is the curvature tensor.
Eq. (\ref{il2.1}) turns out to be a consequence of (\ref{il2.5}) and (\ref{il2.6}) and can be omitted. Eq.
(\ref{il2.5}) determines $e_7=-\frac{R_{(\pi)}}{2M^2}e_1$ while (\ref{il2.6}) gives the lagrangian multiplier
$\lambda_6=\frac{R_{(\pi)}}{2M^2}e_1$. The Dirac-Bergmann algorithm stops at the fourth stage and results in six
non-trivial constraints $T_a$, $a=1, 3, 4, 5, 6, 7$. The multipliers $\lambda_{e1}$, $\lambda_{e3}$ and the variables
$e_1$, $e_3$ have not been determined in the process. This is in agreement with invariance of our action with respect
to two local symmetries mentioned above. Poisson brackets of constraints are shown in Table
\ref{tabular:algebra-constraints-BMT-curved}.

%

%
%
%
%
%
%
\begin{table}
\caption{Algebra of constraints} \label{tabular:algebra-constraints-BMT-curved}
\begin{center}
\begin{tabular}{|c|c|c|c|c|c|c|}
\hline
                              & $\qquad T_1 \qquad$  & $T_3$         & $T_4$          & $T_5$                 & $T_6$    & $T_7$     \\  \hline \hline
$T_1=P^2+m^2c^2 $         & 0             & 0             & 0              & 0                     &$ R_{(\omega)}$   & $R_{(\pi)}$   \\

                              &               &               &           &            &               &             \\
\hline
$T_3=\pi^2-a_3$               & 0             & 0             & $-4T_5$   & $2(T_3-a_3)$          & $-2T_7$ & 0\\
& ${}$ &      &           &                &
&        \\
\hline
$T_4=\omega^2-a_4$            & 0             & $4T_5$  & $0$            & $2(a_4-T_4)$          & 0       & $2T_6$\\
& ${}$ &      &           &                &
&        \\
\hline
$T_5=\omega\pi$               & 0     & $2(a_3-T_3)$      & $2(T_4-a_4)$   &     0                 & $-T_6$  & $T_7$\\
& ${}$ &      &           &                &
&        \\
\hline
$T_6=P\omega$       & $-R_{(\omega)}$  &$2T_7$  & $0$&   $T_6$               & 0       & $-M^2$  \\
                              & ${}$ &      &           &                &
&        \\
\hline
$T_7=P\pi$          & $-R_{(\pi)}$  & 0         & $-2T_6$  & $-T_7$          &$M^2$& 0 \\
                              & ${}$ &       &           &                &
&        \\
\hline
\end{tabular}
\end{center}
\end{table}
The Hamiltonian (\ref{Hamiltonian-curved}) turns out to be a combination of constraints
\begin{equation}\label{Hamiltonian1}
H=\frac{1}{2}e_1\left[ T_1 + \frac{R_{(\pi)}}{M^2} T_6 - \frac{R_{(\omega)}}{M^2} T_7 \right] + \frac{1}{2}e_3\left[ T_3 +
(a_3/a_4)T_4\right]+\lambda_{ei}\pi_{ei}.
\end{equation}
The constraint matrix $\Delta_{ab}\equiv \{ T_a,T_b\}$, which can be easily determinate from the table
\ref{tabular:algebra-constraints-BMT-curved},   has $\mbox{rank}[T_{ab}]=4$. Its null-vectors are $\zeta_1=(M^2,
0,0,0,R_{(\pi)},-R_{(\omega)})$ and $\zeta_2=(0,a_4/a_3,1,0,0,0)$. They determine two first-class constraints of the
model
\begin{eqnarray}
\Psi_1=  T_1 + \frac{R_{(\pi)}}{M^2} T_6 - \frac{R_{(\omega)}}{M^2} T_7, \qquad  \Psi_2= T_3 + (a_3/a_4)T_4.
\end{eqnarray}
As a consequence, the  Hamiltonian  (\ref{Hamiltonian1}) represents a first-class quantity. Adding four
linearly-independent vectors
$\zeta_3=(0,1,0,0,0,0)$, ~$\zeta_4=(0,0,0,1,0,0)$, ~ $\zeta_5=(0,0,0,0,1,0)$ and $\zeta_6=(0,0,0,0,0,1)$
to $\zeta_1$ and $\zeta_2$, we construct an equivalent system of constraints, $\{\Psi_A = \zeta \cdot T\}$. They read
$\Psi_3= T_3$, $\Psi_4=T_5$, $\Psi_5 = T_6$, $\Psi_6 =T_7$. In the set $\{\Psi_A\}$,  constraints $\Psi_1$ and $\Psi_2$
form first-class subset while $\Psi_3,\ldots,\Psi_6$ form a second-class system with non-degenerate constraint-matrix
$\Delta_{ab} = \{\Psi_a , \Psi_b \}$. Its inverse reads
\begin{equation}\label{matrix2}
\Delta^{ab}=\left(
\begin{array}{cccc}
 0& 1/2a_3 & 0 & 0 \\
-1/2a_3 & 0 & 0 & 0 \\
0 & 0 & 0 & 1/M^2  \\
0 & 0 & -1/M^2 & 0  \\
\end{array}
\right).
\end{equation}
This matrix determines Dirac bracket constructed on the base of second-class constraints
\begin{equation}\label{DB}
\{ A ,  B\}_D =\{ A ,  B\}-\{ A ,  \Psi_a\}\Delta^{ab}\{ \Psi_b ,  B\}, \qquad a, b=3, 4, 5, 6.
\end{equation}
Passing to the Dirac bracket, we can work with much more simple Hamiltonian, which does not involve an auxiliary
variables, see Eq. (\ref{H1}) below. We construct the Dirac bracket (see Appendix \ref{DBAppendix}) and note that our
position variables obey highly noncommutative algebra
%
%
\begin{equation}\label{NC}
\{ x^\mu , x^\nu\}_D =\frac{J^{\mu\nu}}{2M^2}\,.
\end{equation}
We can pass from parametric $x^\mu(\tau)$ to physical variables $x^i(t)$. As a consequence of (\ref{NC}), $x^i(t)$ also
obey a noncommutative algebra.

Let us discuss this point in some details. In classical mechanics without constraints, the initial variables usually
have an interpretation as the position variables and obey the Poisson bracket. This situation may change in a theory
with constraints. In a theory with second-class constraints one can find special coordinates on the constraints surface
with canonical (that is Poisson) bracket \cite{gitman1990quantization}. Functions of special coordinates are candidates
for observable quantities. The Dirac bracket (more exactly, its nondegenerated part) is just the canonical bracket
rewritten in terms of initial variables. For the present case, namely the initial coordinates which correspond to
non-degenerated sector (they are $x^i(t)$) are of physical interest, as they represent the position of a particle. So,
while there are special coordinates with canonical symplectic structure, the physically interesting coordinates obey
non-commutative algebra (\ref{NC}). Remind that a theory with second-class constraints can not be consistently
quantized on the base of Poisson bracket. We need to postulate commutators that resemble the Dirac bracket. So the
position variables will be presented by non commutative operators in quantum theory, see \cite{DPM1}.

In summary, to incorporate the Frenkel condition $J^{\mu\nu}P_\mu=0$ in a proper way, we need a model with two
second-class constraints, $P\omega=P\pi=0$, which involve conjugate momenta of $x^\mu$. As a consequence, $x^\mu$ have
nonvanishing Dirac brackets and the position space is endowed, in a natural way, with noncommutative geometry. Our
description of spin on the base of vector variables gives one more example of physically interesting noncommutative
relativistic particle, with the "parameter of noncommutativity" proportional to spin-tensor. There are other examples
where the noncommutative geometry emerges from second-class constraints, see \cite{AAD10, AAD11, Abreu:2010mt,
Amorim:2010qj, Daszkiewicz1310, Daszkiewicz1304, Rivelles1306, Deguchi2012NC, EuneKim2012NC, Horvathy:2003fz}.

We have demonstrated that our spinning particle admits interaction with general curved background, this does not break
neither the number nor the structure of constraints presented in the free theory. Having established the structure of
our model in Hamiltonian formulation, we proceed further analysis at the level of equations of motion.

\section{Hamiltonian and Lagrangian equations of motion in curved background}
The dynamics of basic variables is governed by Hamiltonian equations $\dot z=\{z , H\}$, $z=(x, p, \omega, \pi)$, with
the Hamiltonian (\ref{Hamiltonian1}). After some algebra, these equations can be written in a manifestly covariant form
%
\begin{eqnarray}
\dot x^\mu&=& e_1 \left[ P^\mu +  (2M^2)^{-1}(R_{(\pi)} \omega^\mu-R_{(\omega)}\pi^\mu) \right], \label{motion-x} \\
DP_\mu &=&R^\alpha_{\ \ \beta\nu\mu} \pi_\alpha \omega^\beta  \dot x^\nu, \label{motion-P} \\
D\omega^\mu &=&- e_1\frac{R_{(\omega)}}{2M^2} P^\mu + e_3 \pi^\mu , \qquad D\pi_\mu= -e_1
\frac{R_{(\pi)}}{2M^2} P_\mu - e_3 (a_3/a_4) \omega_\mu. \label{motion-pi-omega}
\end{eqnarray}
Due to presence of arbitrary functions $e_1$ and $e_3$, the evolution of our basic variables is ambiguous. This is in
correspondence with two local symmetries presented in the theory. Position $x^\mu$ and momentum $P_\mu$ have
one-parametric ambiguity due to $e_1$. Basic spin variables $\omega^\mu$ and $\pi_\mu$ have two-parametric ambiguity
due $e_1$ and $e_3$. According to general theory \cite{gitman1990quantization, deriglazov2010classical}, variables with
ambiguous dynamics do not represent observable quantities. So let us look for the variables that can be candidates for
observables. Equation of motion for the spin-tensor
\begin{equation}\label{spin-tensor}
J^{\mu\nu}\equiv 2\left( \omega^\mu \pi^\nu - \omega^\nu \pi^\mu \right) ,
\end{equation}
does not contain $e_3$ (see below). According to geometric construction described in \cite{deriglazov2012variational},
the quantities $J^{\mu\nu}$, being functions of coordinates of the base of spin fiber-bundle, are inert under the spin-plane local symmetry.
By construction, on the constraint surface (\ref{T_5andT_6}) and (\ref{constraints}) the spin-tensor obeys the conditions
\begin{eqnarray}
J^{\mu\nu}P_\nu = 0,\label{condition1} \\
J^{\mu\nu} J_{\mu\nu} = 8a_3a_4. \label{condition2}
\end{eqnarray}
We mention that Eq. (\ref{condition1}) usually imposed by hand in the study of a test-body motion in general relativity
in pole-dipole approximation \cite{Corinaldesi:1951pb}. In our approach this condition is embedded into the model from
the beginning.

Let us write Hamiltonian equations for $x, P$ and $J$. To exclude $\omega$ and $\pi$ from (\ref{motion-x}), we use
(\ref{spin-tensor}) presenting the quantities defined in (\ref{Rpi}) and (\ref{byM.10}) as functions of spin-tensor
\begin{equation}\label{Rpi-Romega-M}
R_{(\pi)}=\frac{1}{2}R_{\alpha\beta\mu\nu}J^{\alpha\beta}P^\mu\pi^\nu,  \quad R_{(\omega)}=\frac{1}{2}R_{\alpha\beta\mu\nu}J^{\alpha\beta}P^\mu \omega^\nu, \quad M^2=m^2c^2-\frac{1}{16}\theta_{\mu\nu}J^{\mu\nu} .
\end{equation}
%
%
%
%
%
Tensor $\theta_{\mu\nu}$ is defined as follows
\begin{equation}\label{byM2}
\theta_{\mu\nu} \equiv R_{\alpha\beta\mu\nu}J^{\alpha\beta}.
\end{equation}
With this, Hamiltonian equation for the coordinate $x^\mu$ can be written in terms of the momentum $P_\mu$ and
spin-tensor $J^{\mu\nu}$. The Eqs. (\ref{spin-tensor}) and (\ref{motion-pi-omega}) immediately imply an equation of
motion for $J^{\mu\nu}$. Using symmetry properties of the curvature tensor, equation (\ref{motion-P}) can be written in
terms of $J$ and $\dot x$. In the result, Hamiltonian equations can be written only in terms of the variables $x$, $P$
and $J$
\begin{eqnarray}
\dot x^\mu &=& e_1\left[ P^\mu -\frac{1}{8M^2}P^\sigma \theta_{\sigma\nu} J^{\nu\mu} \right], \label{motion-x-2.1} \\
DP_\mu &=& \frac{1}{4} R_{\mu\nu\alpha\beta}\dot x^\nu J^{\alpha\beta},  \label{motion-P-2.2} \\
DJ^{\mu\nu} &=&  2\left[P^\mu \dot x^\nu - P^\nu \dot x^\mu \right] \label{motion-J.3}.
\end{eqnarray}
%
%
%
%
%
Together with (\ref{condition1}) and (\ref{condition2}), they form closed system which determines evolution of a
spinning particle. The remaining ambiguity due to $e_1$ is related with reparametrization invariance, and can be
removed in the standard way, passing to the physical-time parametrization. Hence one possible set of candidates for
observables is $(x^i(t), P^i(t), J^{\mu\nu}(t))$.

Hamiltonian can also be rewritten in terms of $x, P$ and $J$. We note that $\pi^2$, $\omega^2$ and $\omega\pi$, being
$SO(1, 3)$\,-invariants, have vanishing brackets with $J^{\mu\nu}$. As a consequence,  the second, third, fourth and
the last terms in (\ref{Hamiltonian-curved}) do not contribute into equations (\ref{motion-x-2.1})-(\ref{motion-J.3}).
So we omit these terms in the Hamiltonian. Further, supposing that we deal with the Dirac bracket, we can use the
constraints $T_6$ and $T_7$ in (\ref{Hamiltonian-curved}). Then the terms proportional to $e_7$ and $\lambda_6$ vanish.
In the result, the Hamiltonian reads
\begin{eqnarray}\label{H1}
H_1=\frac{1}{2} e_1\left(g^{\mu\nu}P_\mu P_\nu + m^2c^2 \right).
\end{eqnarray}
This produces the equations (\ref{motion-x-2.1})-(\ref{motion-J.3}) as follows: $\dot q=\{q, H_1\}_{D}$.

%

%
As it should be expected, they imply second-order equation for the position of a particle. Equations
(\ref{motion-P-2.2}) and (\ref{motion-J.3}) coincide with those of Papapetrou \cite{Papapetrou:1951pa, chicone2005relativistic}.
The r.h.s. of equation (\ref{motion-P-2.2}) points out an influence of spin on the trajectory and can be interpreted as
a spin-orbit interaction. Note that this originates from second term on r.h.s. of the covariant derivative (\ref{s-g}).
In turn, this term is necessary to provide general covariance of the formalism.

Besides, Eq. (\ref{motion-J.3}) implies the equation
\begin{equation}\label{byM12}
D J^{\mu\nu} +\frac{1}{m^2} P^\nu P_\alpha D J^{\mu\alpha}- \frac{1}{m^2}P^\mu P_\alpha D J^{\nu\alpha} = 0,
\end{equation}
which has been taken by Papapetrou as the basic equation for spin, see his equation (5.3).

In the next section we show that (\ref{motion-x-2.1})-(\ref{motion-J.3}) coincide with those of Papapetrou in the
leading-spin limit.

%
%

%
Eliminating $P_\mu$ and $e_1$ from the system (\ref{motion-x-2.1}), (\ref{motion-J.3}), we obtain its Lagrangian form
(for the sector $x$). We first note that the relation between velocity and momentum (\ref{motion-x-2.1}) can be written
as a linear transformation
\begin{equation}\label{motion-x-3}
\dot x^\mu = e_1 T^\mu_{\ \ \nu} P^\nu, \qquad T^\mu_{\ \ \nu} \equiv \delta^\mu_\nu -\frac{1}{8M^2} J^{\mu\alpha} \theta_{\alpha\nu}.
\end{equation}
%
%
%
Using the definition of the spin-tensor (\ref{spin-tensor}), we find the identity
\begin{equation}
J^{\mu\alpha}\theta_{\alpha\sigma}J^{\sigma\beta}\theta_{\beta\nu}=\frac{1}{2} (J\cdot \theta)J^{\mu\alpha}\theta_{\alpha\nu},
\end{equation}
where the term $J\cdot\theta = J^{\mu\nu}\theta_{\mu\nu}$ is  proportional to the gravity-magnetic moment interaction
\cite{Khriplovich96}. The identity allows us to find the inverse matrix
\begin{equation}\label{p-T-x}
(T^{-1})^\mu_{\ \ \nu}= \delta^\mu_\nu + \frac{J^{\mu\alpha}\theta_{\alpha\nu}}{8m^2c^2-J\cdot\theta}.
\end{equation}
Then Eq. (\ref{motion-x-3}) can be inverted,
%
%
and  substituting $P^\mu=P^\mu(\dot x,J, e_1)$ into expression for the constraint $T_1$, $g^{\mu \nu}P_\mu P_\nu +
m^2c^2 =0$, this gives $e_1$
\begin{equation}\label{e_1}
e_1=\frac{\sqrt{-G_{\mu\nu}\dot x^\mu \dot x^\nu}}{mc}\equiv\frac{\sqrt{-\dot xG\dot x}}{mc}.
\end{equation}
We have introduced the effective metric
\begin{equation}
G_{\mu\nu}\equiv (T^{-1})^\alpha_{\ \ \mu}g_{\alpha\beta}(T^{-1})^\beta_{\ \ \nu}.
\end{equation}
The momentum (\ref{momenta-P}), (\ref{p-x}) acquires the following form:
\begin{equation}\label{PTx}
P^\mu = \frac{mc}{\sqrt{-\dot x G \dot x}} ( T^{-1})^\mu_{\ \ \nu}\dot x^\nu =\frac{mc}{\sqrt{-\dot x G \dot x}}\left[ \dot x^\mu + \frac{1}{8m^2c^2-J\cdot\theta} R_{\nu\sigma\alpha\beta}J^{\mu\nu}J^{\alpha\beta}\dot x^\sigma \right].
\end{equation}
Similar expression for $P^\mu$ has been obtained in \cite{Porto} as a compatibility condition of Papapetrou equations
(\ref{motion-P-2.2}) and (\ref{motion-J.3}) with the thansversality condition. Unlike them our momentum $P_\mu$ is
defined through the standard Hamiltonian formalism. Besides, our spin-tensor automatically obeys the Frenkel condition
while in \cite{Porto} the author must demand both (\ref{condition1}) and its conservation by hands.

Substituting (\ref{PTx}) into (\ref{motion-P-2.2}) we get the Lagrangian equation for $x^\mu$
\begin{equation}\label{byM13}
D\left[ \frac{(T^{-1})^\mu_{\ \ \nu} \dot x^\nu}{\sqrt{-\dot xG\dot x}} \right] = \frac{1}{4mc} R^\mu_{\ \ \nu\alpha\beta}\dot x^\nu
J^{\alpha\beta}.
\end{equation}
Similarly, removing $P^\mu$ from (\ref{motion-J.3}) we obtain
\begin{equation}\label{motionJ-5}
DJ^{\mu\nu}= \frac{2mc}{\sqrt{-\dot xG\dot x}}\left[ (T^{-1})^{[\mu}_{\ \ \alpha} \dot x^{\nu]}\dot x^\alpha \right].
\end{equation}
These equations together (\ref{condition1}) and (\ref{condition2}) form closed system for the set $x, J$.

%
\section{First order approximation}
If we study a particle like the electron, then $J^2 \sim\hbar^2$. At the end of the section \ref{subsec:review-BMT} we
mentioned that in the limit $\hbar^2  \ll 1 $ our equations for an electron in uniform electromagnetic field reduce to
those of BMT. In this section we consider a similar approximation, $J^2/c^2\ll 1$, for our particle in gravitational
fields. In such approximation the matrix $T^{-1}$ in (\ref{p-T-x}) reduces to the identity matrix and the effective
metric $G_{\mu\nu}$ coincides with the initial metric tensor. The equation (\ref{byM13}) is simplified
\begin{equation}\label{Motion-5}
D\left[ \frac{\dot x^\mu}{\sqrt{-\dot xg\dot x}} \right] = \frac{1}{4mc} R^\mu_{\ \ \nu\alpha\beta}\dot x^\nu
J^{\alpha\beta}.
\end{equation}
In  (\ref{motionJ-5})   the r.h.s  vanishes
\begin{equation}\label{motionJ-6}
DJ^{\mu\nu}= 0.
\end{equation}
The conditions (\ref{condition1}) and (\ref{condition2})  read as follows: $J^{\mu\nu}\dot x_\nu = 0$, $J^{\mu\nu}
J_{\mu\nu} = 8a_3a_4 \ll c^2$.
%
The equation (\ref{Motion-5}) coincides with the MP equation neglecting the terms nonlinear in spin
\cite{Khriplovich96}. Using the Levi-Civita symbol we define the spin vector
\begin{equation}\label{definition-S}
s^\mu \equiv \frac{g^{-1/2}}{4\sqrt{- \dot xg\dot x}}\varepsilon^{\mu\nu\alpha\beta}\dot x_\nu J_{\alpha\beta} .
\end{equation}
Our approximation implies simple equation for $s^\mu$. To simplify the notation it  is convenient to chose  the
proper-time parametrization $s$. The equation for the trajectory of the test particle now  reads
\begin{equation}\label{Motion-6}
D_su_\mu = \frac{g^{-1/2}}{2mc} R_{\mu\nu\alpha\beta}\varepsilon^{\alpha\beta\sigma\kappa}s_\sigma u_\kappa u^\nu,
\end{equation}
where $u_\mu\equiv (dx_\mu/ds)$. Defining
\begin{equation}
F_{\mu\nu} \equiv \frac{g^{-1/2}}{2}R_{\mu\nu\alpha\beta}\varepsilon^{\alpha\beta\sigma\kappa}s_\sigma u_\kappa,
\end{equation}
equation (\ref{Motion-6}) can be written in terms of effective Lorentz force
\begin{equation} \label{motion-u}
D_su_\mu = \frac{1}{mc}F_{\mu\nu} u^\nu.
\end{equation}
%
%
%
%
Equation (\ref{motionJ-5}) and definition (\ref{definition-S}) imply that
\begin{equation}\label{motionS-2}
D_ss^\mu= 0,
\end{equation}
where we have taken into account that $F_{\mu\nu}$ is of first order in $S^\mu$ ($\sim \hbar$ for the electron). The
supplementary spin conditions for $s^\mu$ are
\begin{equation}
s^\mu u_\mu = 0, \quad s^\mu s_\mu=a_3a_4\ll c^2 . \label{conditionS-2}
\end{equation}
Equations (\ref{motion-u}), (\ref{motionS-2}) and  (\ref{conditionS-2}) determine behavior of both trajectory and spin
in a gravitational field.

Let us consider  the weak-field approximation
\begin{equation}
g_{\mu\nu}=\eta_{\mu\nu}+h_{\mu\nu}, \quad |h_{\mu\nu}|\ll1,
\end{equation}
and a spherically symmetric field produced by a mass $M_s$
\begin{equation}
h_{00}=\frac{2kM_s}{c^2 r}, \quad h_{ij}= -\frac{2kM_s}{c^2r}\delta_{ij}, \quad h_{0i}=0.
\end{equation}
The corresponding equation of motion for a test spinning particle derived from Eq. (\ref{Motion-5}) (or, alternatively,
form (\ref{motion-u})) is
\begin{equation}\label{motion-r}
{\bf\ddot x} = -\frac{kM_s}{c^2}\frac{{\bf x}}{x^3}  - 3\frac{kM_s}{mc^2}\frac{1}{x^3}\left[ [{\bf\dot x}\times {\bf
S}]-({\bf \hat n\cdot\dot x})[{\bf \hat n \times S}]  -  2([{\bf\dot x}\times {\bf S}]{\bf \cdot \hat n) \hat n
}\right].
\end{equation}
Here ${\bf \hat n = x/}x$ and ${\bf S}$ is the spacial part of spin vector (\ref{definition-S}), $M_s$ is the mass of
the source of the gravitational field and $m$ is the mass of the spinning test particle. The equation coincides with
the MP equation taken in the weak field approximation, see \cite{Barker}. Hence, in all practically interesting cases,
evolution of the Frenkel particle in a gravitational background coincides with that of a rotating body in pole-dipole
approximation.

The acceleration (\ref{motion-r}) is different from that obtained in the quantum theory of Dirac equation
\cite{Barker70}
\begin{equation}\label{motion-Rr}
{\bf\ddot r} = -\frac{kM_s}{c^2}\frac{{\bf r}}{r^3} - 3\frac{kM_s}{mc^2}\frac{1}{r^3}\left( [{\bf\dot r}\times {\bf
S}]-\frac{3}{2}({\bf  n\cdot\dot r})[{\bf  n \times S}]  -  \frac{3}{2}{\bf( [\dot r\times S]\cdot  n)  n }\right),
\end{equation}
where ${\bf n=r}/r$. The last equation also can be obtained if one starts with classical mechanics of two spinning
particles with spin-orbit interaction \cite{Khriplovich96}.

One possible explanation of the difference between (\ref{motion-r}) and (\ref{motion-Rr}) has been suggested in
\cite{Pomeranskii, Khriplovich96, Barker}. For the case of a rotating object, one can adopt different definitions of
its center-of-mass. If we relate the variables ${\bf x}$ and ${\bf r}$ (in the considered approximation) as follows:
\begin{equation}\label{shift}
{\bf x}={\bf r} + \frac{1}{mc}\left[ {\bf \dot r \times S}\right]\,,
\end{equation}
the equation (\ref{motion-r}) turn into (\ref{motion-Rr}). So the discrepancy may be explained as coming from different
definitions of the center of mass for a rotating object.

Let us point out another explanation. In a number of classical-mechanical models of Dirac equation \cite{AAD5, AAD12, AAD6,
AAD3}, the variable ${\bf r}$, which corresponds to the position operator $\hat{\bf r}$ of the Dirac equation,
is a gauge non-invariant variable. So ${\bf r}$ is non-observable quantity. The observable quantity ${\bf x}$ turns out
to be related with ${\bf r}$ just according the equation (\ref{shift}).

Concerning the equation (\ref{shift}), we also point out that in the approximation under consideration, the Dirac
brackets (\ref{DB}) imply $\{ r^i , r^j\}_D = 0$, that is the radius-vector in Eq. (\ref{motion-Rr}) corresponds to the
canonical coordinates of our theory, see discussion at the end of section \ref{sec:curved-ST}.

%

%


\section{Conclusions}

In this work we have extended the variational formulation of a spinning particle in an electromagnetic field \cite{Alexei} to
the case of a curved space-time background. This demonstrates an effectiveness and generality of the classical description
of spin on the base of vector-like variable. Our model allows interaction with an arbitrary curved background without
modifications of the number and the structure of constraints (\ref{intr.22})-(\ref{intr.21}). The supplementary spin
conditions (\ref{condition1}), (\ref{condition2}) are guaranteed by the constraints (\ref{intr.20}) and (\ref{intr.21})
arising from our singular Lagrangian (\ref{L-curved}).

We derived a closed system of Hamilton equations (\ref{motion-x-2.1})-(\ref{motion-J.3}) for the position,  momentum and
spin-tensor. If we work with Dirac brackets, these equations can be obtained as Hamiltonian equations with a simple
Hamiltonian written in Eq. (\ref{H1}). As expected, position coordinates obeys the second-order equation, while the
equations (\ref{motion-P-2.2}) and (\ref{motion-J.3}) are compatible with those of Papapetrou \cite{Papapetrou:1951pa}.
In the practically important case of the leading-spin approximation our equations coincide with those of Papapetrou.

Excluding the momenta $P^\mu$ from the Hamiltonian equations, we obtained their Lagrangian form in terms of $x^\mu$ and
$J^{\mu\nu}$, see equations (\ref{byM13}) and (\ref{motionJ-5}). Linear in spin approximation of these equations yields
a simple equation for the spin-vector $S^\mu$. Combining this with the weak-field approximation used in general
relativity, we get equations of motion for a spinning object in a gravitational field with spherical symmetry. We hope
that the developed variational model will be useful for the description and analysis of experimental setups involving
relativistic spinning particles in a gravitational background.

It is known that the formalism of dynamical systems with second-class constraints implies a natural possibility to
incorporate noncommutative geometry into the framework of classical and quantum theory \cite{AAD10, AAD11}. Our model
represents an example of a situation in which physically interesting noncommutative particle emerges in this way. In our
formalism, components of the position variable have non-trivial Dirac brackets originating from the necessity to take
into account the spin-tensor condition $J^{\mu\nu}\dot x_\mu=0$. In the result, we deal with a noncommutative position
coordinate instead of the canonical one. The description of spin on the base of vector-like variables leads inevitably
to the position noncommutativity, Eq. (\ref{NC}).

In general relativity,  a consistent definition of quantum spin is a nontrivial task
\cite{singh2007cov-quantum-spin}. We note that with the variational formulation at hand, quantization of a theory
represents a more or less straightforward procedure, thus opening the way to further studies of microscopic phenomena
involving spin, gravity and inertia \cite{mashhoon2000}. For instance, accounting for spin-rotation coupling
\cite{Lambiase2013spin-rotation-coupling} is important for precise measurements of anomalous magnetic momentum of
elementary particles \cite{papini2002spin}.

\appendix
\section{Dirac brackets}\label{DBAppendix}
We construct Dirac brackets that take into account the subset of constraints $\{\Psi_a\}$ with non degenerated second-class constraint matrix (\ref{matrix2}). The Dirac bracket between any two phase space functions $A$ and $B$ is given by (\ref{DB}). For the basic variables we obtain
\begin{eqnarray}
\{ x^\mu, x^\nu \}_D &=& \frac{J^{\mu\nu}}{2M^2}, \\
\{ x^\mu, P_\nu \}_D &=&\delta^\mu_\nu -\frac{1}{8M}\left[\theta_{\alpha\nu}J^{\mu\alpha} +4\Gamma^\alpha_{\nu\beta}P_\alpha J^{\beta\mu} \right], \\
\{ x^\mu,\omega^\nu \}_D &=& \frac{\omega^\mu P^\nu}{M^2}+\frac{1}{2M^2}\Gamma^\nu_{\alpha\beta}\omega^\beta J^{\alpha\mu}, \\
\{ x^\mu, \pi^\nu \}_D &=&\frac{\pi^\mu P^\nu}{M^2}-\frac{1}{2M^2}\Gamma^\nu_{\alpha\beta}\pi^\alpha J^{\mu\beta}, \\
\{ x^\mu,J^{\alpha\beta} \}_D &=& \frac{P^{[\beta} J^{\alpha]\mu}}{M^2}+\frac{1}{2M^2}J^{\mu\sigma} \Gamma_{\sigma\nu}^{[\alpha} J^{\beta]\nu},  \\
\{ P_\mu, P_\nu\}_D &=& \frac{1}{4}\theta_{\mu\nu} - \frac{\theta_{\alpha\mu}  \theta_{\beta\nu}  J^{\beta\alpha} }{32M^2} + \\
&&\frac{1}{4M^2}\left[ \theta_{\alpha\mu} \Gamma_{\nu\beta}^\lambda  P_\lambda  J^{\alpha \beta} + \theta_{\beta\nu}  \Gamma_{\mu\alpha}^\lambda  P_\lambda  J^{\alpha\beta} +4 \Gamma^{\alpha}_{\mu\beta}  \Gamma^\lambda_{\nu\sigma}  P_\alpha  P_\lambda J^{\sigma\beta}  \right], \nonumber \\
\{ P_\mu, \omega^\nu \}_D &=&\Gamma^\nu_{\mu\alpha}  \omega^\alpha - \frac{1}{4}\left[\theta_{\alpha\mu}  \omega^\alpha  P^\nu +\frac{1}{2} \theta_{\alpha\mu}  \Gamma^\nu_{\sigma\lambda}  \omega^\lambda  J^{\sigma\alpha} - 4 \Gamma^{\alpha}_{\mu\beta} \Gamma^\nu_{\sigma\lambda} P_\alpha \pi^\beta  \omega^\sigma   \omega^\lambda    \right] , \\
\{P_\mu , \pi_\nu \}_D &=&- \Gamma^\alpha_{\mu\nu}\pi_\alpha -\frac{1}{4}\left[\theta_{\alpha\mu}  \pi^\alpha  P_\nu +\frac{1}{2}\theta_{\alpha\mu}  \Gamma^\sigma_{\nu\lambda}  \pi_\sigma J^{\alpha\lambda} - 4\Gamma^\alpha_{\mu\beta} \Gamma^\sigma_{\nu\lambda}  P_\alpha  \omega^\beta  \pi^\lambda  \pi_\sigma \right], \\
\{ \omega^\mu , \pi_\nu \}_D &=& \delta^\mu_\nu -\frac{\pi^\mu\pi_\nu}{a_3} + \nonumber \\
&& \frac{1}{M^2}\left[ \Gamma^\mu_{\alpha\beta} \omega^\alpha  \omega^\beta \Gamma^\sigma_{\nu\lambda} \pi_\sigma  \pi^\lambda   +  ( P^\mu  - \Gamma^\mu_{\alpha\beta}  \pi^\alpha  \omega^\beta ) ( P_\nu - \Gamma^\sigma_{\nu\lambda} \omega^\lambda  \pi_\sigma \right].
\end{eqnarray}
The tensor $\theta_{\mu\nu}$ is defined in (\ref{byM2}).

\acknowledgments

The work of AAD has been supported by the Brazilian foundation CNPq. AMPM and WGR
thanks CAPES for the financial support (Programm PNPD/2011).



\end{document}